\documentclass{bmcart}

\usepackage{amsthm,amsmath}
\usepackage{url}
\usepackage{graphics}
\usepackage[utf8]{inputenc} 
\usepackage{graphicx}
\usepackage{xfrac}
\usepackage{amsfonts}
\usepackage{amssymb}
\usepackage{float}
\usepackage{color}
\usepackage{makecell}
\definecolor{womencolor}{RGB}{117,112,179}
\definecolor{mencolor}{RGB}{27,158,119}
\definecolor{unknowncolor}{RGB}{217,95,2}
\newcommand{\change}[1]{\textcolor{black}{#1}}
\newcommand{\minorchange}[1]{\textcolor{black}{#1}}
\newcommand{\newchange}[1]{\textcolor{black}{#1}}
\newcommand{\changelastround}[1]{\textcolor{black}{#1}}

\newcommand{\spaceaftersection}{\vspace{10pt}}
\newcommand{\spaceaftersubsection}{\vspace{7pt}}

\makeatletter
\newcommand*{\rom}[1]{\expandafter\@slowromancap\romannumeral #1@}
\makeatother

\usepackage{tabu}
\usepackage{array, makecell}
\usepackage{boldline}

\startlocaldefs
\endlocaldefs

\begin{document}

\begin{frontmatter}

\begin{fmbox}
\dochead{Research}

\title{Gender and Collaboration Patterns\\ in a Temporal Scientific Authorship Network}

\author[
   addressref={aff1},                   
   corref={aff1},                       
   email={geciah@princeton.edu}   
]{\inits{GBH}\fnm{Gecia} \snm{Bravo-Hermsdorff}}
\author[
   addressref={aff1,aff2},
   email={valkyrie.felso@gmail.com}
]{\inits{VF} \fnm{Valkyrie} \snm{Felso}}
\author[
   addressref={aff3,aff4},
   email={emilyremilyray@gmail.com}
]{\inits{ER}\fnm{Emily} \snm{Ray}}
\author[
   addressref={aff5},
   email={leeg@princeton.edu}
]{\inits{LMG}\fnm{Lee M.} \snm{Gunderson}}
\author[
   addressref={aff6},
   email={meheland@syr.edu}
]{\inits{MEH}\fnm{Mary E.} \snm{Helander}}
\author[
   addressref={aff4},
   email={joanamaria@us.ibm.com}
]{\inits{JM}\fnm{Joana} \snm{Maria}}
\author[
   addressref={aff1,aff2},
   email={yael@princeton.edu}
]{\inits{YN}\fnm{Yael} \snm{Niv}}


\address[id=aff1]{
  \orgname{Princeton Neuroscience Institute, Princeton University}, 
  \street{Washington Rd},                     %
  \postcode{08544},                                
  \city{Princeton},                              
  \cny{NJ, USA}                                    
}
\address[id=aff2]{
  \orgname{Department of Psychology, Princeton University}, 
  \street{Washington Rd},                     %
  \postcode{08544},                                
  \city{Princeton},                              
  \cny{NJ, USA}                                    
}
\address[id=aff3]{%
  \orgname{Product Data Science, Grubhub Inc.},
  \street{1065 6th Ave},
  \postcode{10018},
  \city{New York},
  \cny{NY, USA}
}
\address[id=aff4]{%
  \orgname{Department of Data Science, IBM T.J.~Watson Research Center},
  \street{1101 Kitchawan Rd},
  \postcode{10598}
  \city{Yorktown Heights},
  \cny{NJ, USA}
}
\address[id=aff5]{%
  \orgname{Princeton Plasma Physics Laboratory, Princeton University},
  \street{100 Stellarator Rd},
  \postcode{08540}
  \city{Princeton},
  \cny{NJ, USA}
}
\address[id=aff6]{%
  \orgname{Maxwell School of Citizenship and Public Affairs, Syracuse University},
  \street{426 Eggers Hall},
  \postcode{13244},
  \city{Syracuse},
  \cny{NY, USA}
}

\begin{artnotes}
\end{artnotes}

\end{fmbox}

\begin{abstractbox}
\begin{abstract} 
One can point to a variety of historical milestones for gender equality in \minorchange{STEM (science, technology, engineering, and mathematics)}, however, the practical effects are gradual and ongoing. 
It is important to quantify gender differences in subdomains of scientific work in order to detect potential \minorchange{biases} and to monitor progress.  
In this work, we studied the relevance of gender in scientific collaboration patterns in the Institute for Operations Research and the Management Sciences (INFORMS), a professional organization with sixteen peer-reviewed journals. 
We constructed a large \minorchange{temporal bipartite network between authors and publications,} using the organization's publication data from 1952 to 2016, and augmented the author nodes with gender labels.
We characterized differences in several basic statistics of this network over time, highlighting \minorchange{how they change} with respect to relevant historical events. 
We found a steady increase in participation by women (e.g., fraction of authorships by women \minorchange{and} of new women authors) starting $\sim$1980\minorchange{. 
However,} women still comprise less than 25$\%$ of the INFORMS society\minorchange{, and are additionally underrepresented among authors with many publications}.
Finally, we describe a methodology for quantifying differences in the role \minorchange{that} authorships by women and men \minorchange{play} in the overall connectivity of the network. 
Specifically, we propose a degree-preserving temporal and geometric null model with emergent communities\minorchange{.  
We use two measures of edge importance related to diffusion throughout the network, namely} effective resistance and edge contraction \minorchange{importance to} quantify gender differences in collaboration patterns \minorchange{that go} beyond differences in local statistics. 
\end{abstract}


\end{abstractbox}

\end{frontmatter}

\section*{Introduction}
\spaceaftersection

Recent years have seen increasing awareness and discussion of systematic gender biases in academia. 
A slew of studies and opinion publications~\cite{raymond2013most, schrouff2019gender} have highlighted often unintentional, but nevertheless pervasive, biases in integral aspects of academic careers such as hiring practices~\cite{mossracusin2012science}, funding decisions~\cite{witteman2019gender}, peer review~\cite{helmer2017gender}, and representation of women as speakers in conferences~\cite{nittrouer2017gender}.  
It is therefore important to precisely quantify gender differences in subdomains of academic work as a first step towards detecting potential biases and finding appropriate solutions. 

One domain in which network science could offer insight is in the study of gender differences in authorship and collaboration patterns. 
Indeed, there have been several studies on this topic.
For example, Ara\'ujo et al.~\cite{araujo2017gender} analyzed \minorchange{a dataset} containing more than 270,000 academics from a wide variety of fields \minorchange{(engineering,} arts, biological, exact, and social sciences) in Brazil. 
They found that, across all fields, men were more likely to collaborate with other men, while the collaboration gender ratios of women more closely matched that of the relevant academic population. 
In contrast, Karimi et al.~\cite{karimi2018analyzing} showed that, in \minorchange{a} research community \minorchange{in organizational science} based in the US and Europe, women exhibited more gender \minorchange{homophily.} 
\change{Jadidi et al.~\cite{jadidi2018gender} investigated gender differences in a temporal network spanning 47 years with publication and citation data of over one million computer scientists, concluding that homophily has been increasing recently for both genders. 
Additionally, they found that women have a higher dropout rate than men, especially at the beginning of their careers. 
West} et al.~\cite{west2013therole} studied gender differences in a corpus containing over eight million publications across a variety of fields \minorchange{(humanities, social, and natural sciences)}. They found that, overall, gender differences in number of publications have been decreasing over time.  However, in some fields, women remain disproportionately underrepresented as first, last, and \minorchange{solo} authors.

While these studies provide a glimpse into gender differences in authorship and collaboration networks in academia, they tend to focus on local statistics (i.e., measures that depend only on neighboring connections, such as number of publications, homophily, and author order). 
\minorchange{In contrast,} relatively less is known about the correlation between author gender and their roles in the global structure of the network.  
\minorchange{Indeed, appropriately} characterizing these differences \minorchange{is} a nontrivial task, as local and global measures are often intertwined.

\minorchange{In this work, we investigate gender differences over time using both local and global measures in} a large corpus of publications from 1952 to 2016 in journals affiliated with the Institute for Operations Research and the Management Science (INFORMS) -- the predominant professional society for the disciplines of operations research and management science~\cite{aboutinforms}. These fields are both squarely within STEM (science, technology, engineering, and \minorchange{mathematics}), where the degree of gender imbalance has traditionally been considered more severe (but see~\cite{leslie2015expectations}). 

We represent the data as a temporal bipartite \minorchange{(between authors and publications)} network, and first characterize several ``local'' \minorchange{statistics} of this network, highlighting their change with respect to potentially relevant historical events. 
We then describe a methodology for quantifying differences with respect to ``global'' structure (\minorchange{conditioned on several relevant} local statistics)\minorchange{, which} could be of independent \minorchange{interest.}  

\change{Specifically, we consider ``global'' structure to be related to the prototypical global process of diffusion.  
To this end, we selected two measures related to such dynamics, namely, effective resistance~\cite{spielman2011graph,chandra1989theelectrical,christiano2011electrical} and edge contraction importance~\cite{bravohermsdorff2019aunifying}, to measure the global importance of an author--publication connection.  }
\minorchange{To account for local statistics,} we describe a simple temporal and geometric null model, with only \minorchange{two} free parameters that control the emergence and size of communities. 
\minorchange{This} null model explicitly replicates \minorchange{the observed yearly degree distributions for publications, as well as the genders and publication histories of each individual author. 
Importantly,} the \minorchange{mechanism it} uses to decide which authors participate in which publications \minorchange{is blind to the author} gender.  
By comparing the null model to the \minorchange{observed} network data, we can identify \mbox{gender-related} differences in the structural role \minorchange{played by men and women authorships that go beyond} local \minorchange{statistics.
Given} the increased interest in the relatively new field of fairness in machine learning~\cite{liu2018delayed, chouldechova2018frontiers}, the type of graph-theoretical analysis that we propose here could be useful for bias detection in other social networks. 


\bigskip

\section*{The dataset: a temporal, bipartite network\\ \minorchange{between authors and their publications in INFORMS}}
\spaceaftersubsection

In this section, we introduce some relevant information about the INFORMS society, and describe our data acquisition and \minorchange{cleaning} methods.

\subsection*{\textbf{The INFORMS society}}
\spaceaftersubsection

INFORMS was founded in 1995 with the merging of \minorchange{two societies}: the Operations Research Society of America (ORSA) and The Institute of Management Sciences (TIMS) \cite{informs_history}\minorchange{, two societies that had already been linked prior to this time.}  
\minorchange{For instance, they had many members in common \cite{1973orsatims}, had hosted joint meetings since 1961 \cite{hall1983issue}, and jointly sponsored two journals (\textit{Mathematics of Operations Research} \cite{moor_found}, founded in 1976, and \textit{Marketing Science} \cite{mksc_found}, founded in 1982).}
\mbox{INFORMS} currently publishes 16 peer-reviewed journals \cite{informs_journals}, the oldest of which are the flagship journals of the two original societies: \textit{Operations Research}, first published in 1952 by ORSA and \textit{Management Science}, first published in 1954 by TIMS. 
\minorchange{As of December 2017, the} society reports to have over 12,500 members \cite{aboutinforms}, around 20\% of which identify as women  \cite{presidentsdesk}. 

The INFORMS society has historically worked towards identifying and mitigating gender bias, making it particularly attractive for our study. For example, in 2006, then-president Mark S.~Daskin founded a diversity committee ``to assess whether or not there is any sort of problem with diversity within INFORMS'' \cite{nagurney}.  
In 2017, the society created a ``Diversity, Equity, and Inclusion'' initiative, including the formation of a committee \minorchange{with} 
\begin{quote}
``[...] a broad charge that includes monitoring the diversity of our membership and seeking out, creating and maintaining best practices for INFORMS to improve diversity and inclusion-related performance.'' \cite{presidentsdesk}  
\end{quote}
\noindent

\subsection*{\textbf{Data acquisition, data \minorchange{cleaning}, and gender assignments}} 
\spaceaftersubsection

We constructed a bipartite authorship network \minorchange{using the} publications from 16 peer-reviewed journals affiliated with INFORMS from 1952 to 2016.\footnote{\footnotesize{The 16 journals in our dataset are not identical to the 16 journals currently published by INFORMS as one journal was removed during our period of interest (\minorchange{\textit{Management Technology}} merged with \minorchange{\textit{Management Science}} in 1965), and one was added after our period of interest (INFORMS introduced a new journal, \minorchange{\textit{INFORMS Journal on Optimization}} in 2018).}} Publication records were acquired using INFORMS PubsOnline (in the form of BibTeX entries) and the CrossRef REST API \cite{informspubonline, crossref}. 
Nodes in the network are of two types: author or \minorchange{publication, and edges} denote an authorship of a publication. 
Metadata include author name, \minorchange{author order,} publication title, year, and journal.

\minorchange{We classified authors by gender using the commercial package Genderize.io API \cite{genderizeio}, which associates to each first name a gender (woman or man) and a confidence score ranging from 0 to 100. 
We thresholded} acceptable gender labels as those with a confidence level above 80, noting that this could still yield misclassifications. 
Author nodes for whom \minorchange{the} confidence was below 80 were initially marked as ``unknown'' gender. 

We performed two data \minorchange{cleaning} steps: (1) additional gender classification and (2) author node combination. 
For (1)\change{, all unknown-labeled authors with more than 6 publications were manually classified via internet searches, and based on our personal knowledge of authors in the field. 
In addition, we verified the gender of the $100$ most prolific woman-labeled authors, as well as the $100$ most prolific men-labeled authors}, and corrected the labels for any who had been \minorchange{misclassified}. 
These \minorchange{misclassifications} had occurred mostly for given names that were historically used for men and shifted to women over time (e.g., Aubrey, Leslie, Sandy).
For (2), we combined multiple nodes that referenced the same author (usually due to publication under a variety of names). For example, Robert Eugene Donald Woolsey, a prolific and well-known figure in the INFORMS \minorchange{society}, published under the names ``Eugene Woolsey'', ``Gene Woolsey'', ``R.~E.~D.~Woolsey'', ``R.~E.~Woolsey'', and ``Robert~E.~D.~Woolsey''. 
We manually verified the $100$ most prolific author nodes (along with any others noticed while examining the data) by searching for duplicate last (family) names and combining authors as needed.

The resulting network was composed of $23875$ publication nodes, $22911$ author nodes, $50527$ edges, and $4587$ connected components.  
Of the author nodes, $16179$ were labeled as men, $2997$ as women, and $3735$ as unknown. 
The giant (i.e., largest) component of the network contained $13520$ author nodes, $16604$ publication nodes, and $38169$ edges.   
\change{Often, in studies of collaboration networks, one ``projects'' the network to a single type of node (e.g., by keeping the author nodes and replacing the publication nodes with cliques of edges between those authors)~\cite{araujo2017gender, karimi2018analyzing, liu2005coauthorship, abbasi2012eocentric, chen2017building, jadidi2018gender}.  
However, we kept the original bipartite structure in all of our analysis as projecting the network loses 
structural information~\cite{kitsak2017latent} 
(e.g, a publication with three authors would be identical to three publications between all pairs of those authors).
We also kept the connections by authors of unknown gender to maintain the overall network connectivity.}

\section*{\minorchange{Local statistics}: participation rates by women are increasing,\\ but remain far from gender parity}
\spaceaftersection

The broad strokes of gender asymmetry are easily seen with simple measures, such as participation rates. 
In this section, we \minorchange{quantify and} discuss the evolution of several local statistics in the INFORMS authorship network.
\change{Additional local statistics are contained in Supplementary Information Table~\ref{table:summarystatistics}.}

\subsection*{\textbf{Relevant historical milestones}}
\spaceaftersubsection

While gender discrimination continues to be an issue in academic environments, actions have been taken to mitigate this discrepancy. 
To place the data in the relevant historical context, we highlight in our graphs the timing of two notable examples of such events:\footnote{\footnotesize{Both actions we mention took place in the United States, where $\sim$70\% of \mbox{INFORMS} members are currently based~\cite{magrogan}.}}
\begin{enumerate}
\item \textit{Title IX of the Education Amendments Act of 1972,} which states:
\begin{quote}
No person in the United States shall, on the basis of sex, be excluded from participation in, be denied the benefits of, or be subjected to discrimination under any education program or activity receiving Federal financial assistance~\cite{titleix}.
\end{quote}
While Title IX focused primarily on student athletics, the policy has had far-reaching consequences. 
For example, the law has been credited with increasing access to college education for women~\cite{titleixreport} \minorchange{ and supporting} applications to further gender equality for those employed in academia \cite{walters2010recasting}.  
 \item \textit{The Family and Medical Leave Act of 1993 (FMLA)}~\cite{fmla}.  
This law protects employees who take leave from work for family or medical reasons.
As women have traditionally performed most of the child care in the family, it represents another major milestone for women's rights in the United States, where family leave (including for childbirth) was not otherwise available or protected by law~\cite{klerman2012family}.
\end{enumerate}

\subsection*{\textbf{Participation rates by women and men over time}}
\spaceaftersubsection

\begin{figure}[ht]
\includegraphics[width=0.7\linewidth]{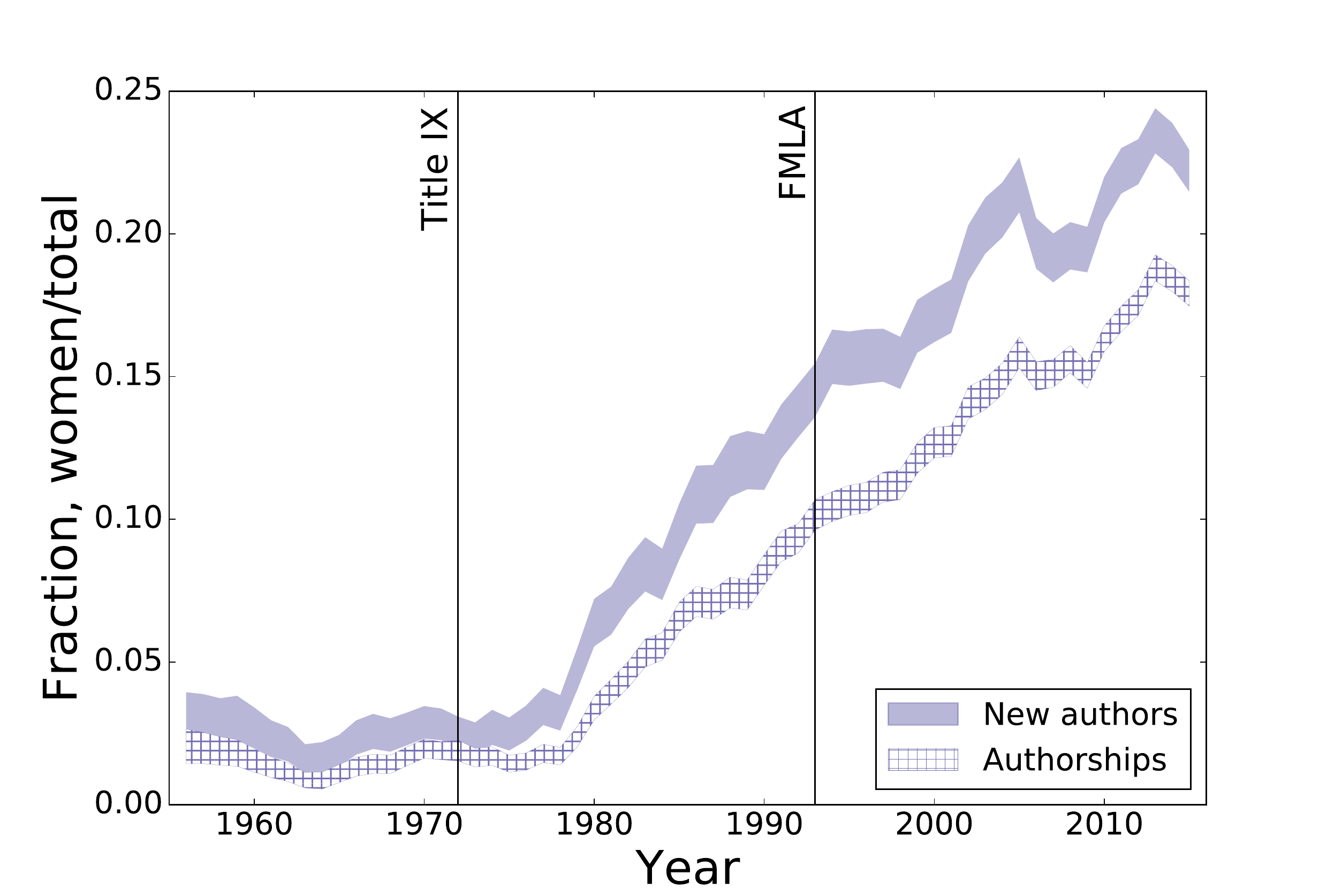}%
\caption{\csentence{Participation rate by women in INFORMS has been increasing since $\sim$1980, but remains well below gender parity.}
  \textit{Hatched:}~Fraction of authorships by women over time.
  \textit{Solid:}~Fraction of new women authors over time. 
The data were binned in a moving window centered around the plotted year. 
To allow for better statistics\minorchange{, for} 1956--1963 \minorchange{the window widths were 8 years,} \minorchange{for} 1964--1971 \minorchange{they were} 6 years, and the remainder \minorchange{were} 4 years.
Fractions were calculated excluding authors without a gender label.
Shading denotes $\pm1$ standard error of the mean, estimated treating the \minorchange{gender labels of all datapoints in the relevant window} 
as a collection of \minorchange{i.i.d.}~Bernoulli random variables (i.e., $p \pm \sqrt{p (1-p) / n}$, where $p$ is the fraction and $n$ is the number of datapoints in the \minorchange{window}).  
Note that due to the binning the rise in the curves in \minorchange{$\sim$}1978 reflects an increase of participation by women starting in \minorchange{$\sim$}1980. 
   }
\label{fig:authorship}  
\end{figure}

The fractions of new women authors and of authorships by women have both been increasing since $\sim$1980 (Figure~\ref{fig:authorship}). 
However, both measures remain more than a factor of 2 from gender parity.
In fact, a crude extrapolation of the fraction of new women authors (a weighted linear regression from 1976--2016) yields an estimate of gender parity by 2062$\pm$5.
The same extrapolation for \minorchange{the} fraction of women authorships gives an estimate of gender parity by 2083$\pm$3 (standard deviations were computed by propagating the errors due to the covariance of the slope and intercept of the linear fit).

\begin{figure}[h!]
\includegraphics[width=0.98\linewidth]{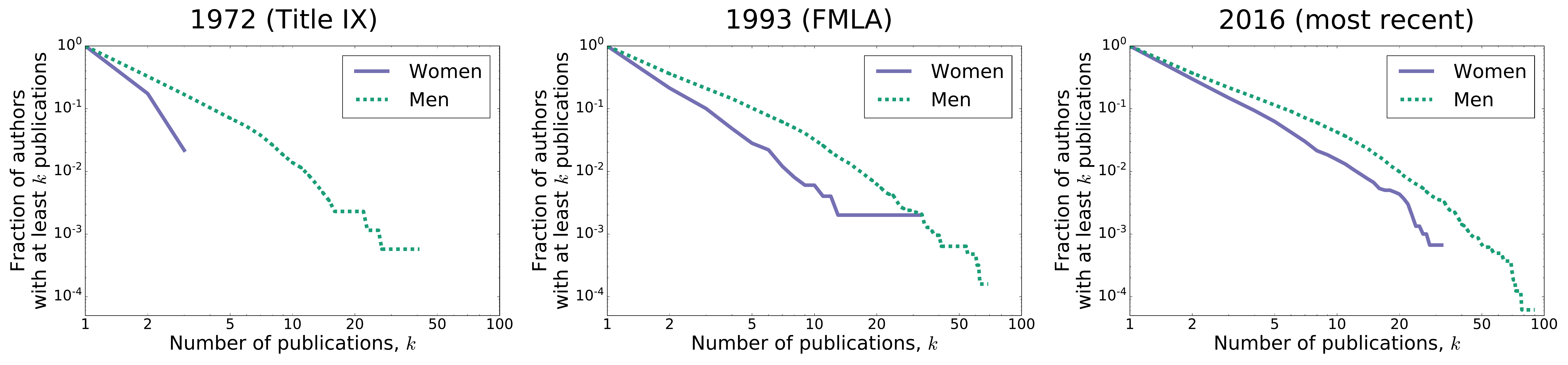}%
\caption{
\minorchange{\csentence{Evolution of the cumulative degree distributions in the INFORMS network \minorchange{suggests} that women are disproportionately underrepresented among authors with many publications.
}}
Plots display the normalized cumulative author degree distribution \minorchange{for the indicated year, }i.e., the vertical axis is the fraction of authors of this gender with at least that number of \minorchange{publications.} 
Since there are more men than women in the network, the degree distribution for men is expected to extend to more publications than that for women. 
However, the steeper slope of the distribution for women (in log scale) suggests \minorchange{their} systematic underrepresentation, especially in the high-publication tail of the distribution \minorchange{(quantified in Figure~\ref{fig:powerlaw}).}   
See supplementary material for a video of the cumulative degree distributions \minorchange{over time.}  
}
\label{fig:degreedistribution}  
\end{figure}

\begin{figure}[h!]
\includegraphics[width=0.97\linewidth]{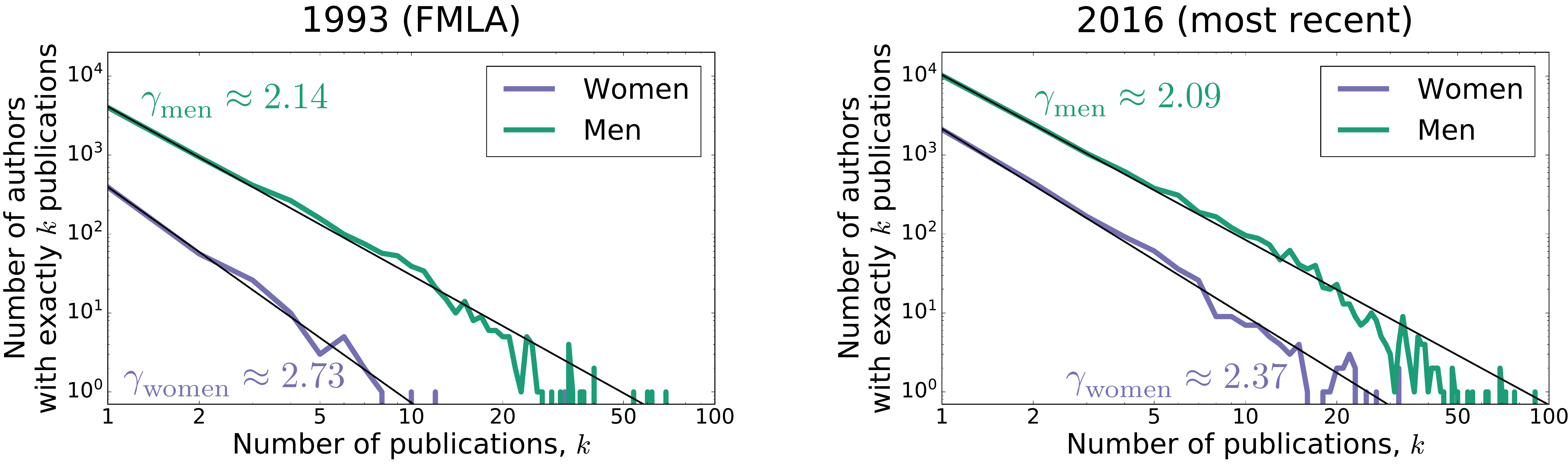}
\caption{
\change{\csentence{Women are disproportionately underrepresented among authors with many publications in INFORMS.}} 
There appears to be an additional underrepresentation of women among authors with many publications, beyond what would already be expected by the lower proportion of women authors in the data. 
To quantify this effect, we modeled the degree distributions of each gender as a power law, using the exponent $\gamma$ as a measure of how ``heavy'' the tail of these distributions are.  
By considering the number of authors with $k$ publications as an independent Poisson variable with mean $n(k) = c k^{-\gamma}$, we maximized the likelihood of the data over the space of $c$ and $\gamma$. 
The results showed a steeper power law for women ($\gamma_{\text{women}} \approx 2.73$ vs.~$\gamma_{\text{men}} \approx 2.14$ in 1993 and $\gamma_{\text{women}} \approx 2.37$ vs.~$\gamma_{\text{men}} \approx 2.09$ in 2016). 
To quantify the significance of this difference, we repeatedly randomized the gender labels of the nodes and considered the distributions of the fitted exponents $\gamma$ as the null distribution for the observed result.  
We found that the \mbox{$z$-scores} of the observed \mbox{power-law} exponents were about $- 9.4$ for men and $+11.6$ for women in 1993, and about $- 6.9$ for men and $+10.5$ for women in 2016, all highly significant.  
\changelastround{This indicates that women are disproportionately underrepresented among authors with many publications in INFORMS.}
However, this difference seems to be attenuating, as suggested by the decrease in the difference between $\gamma_{\text{women}}$ and $\gamma_{\text{men}}$ during this time. 
}
\label{fig:powerlaw}
\end{figure}

\begin{figure}[ht]
\includegraphics[width=0.97\linewidth]{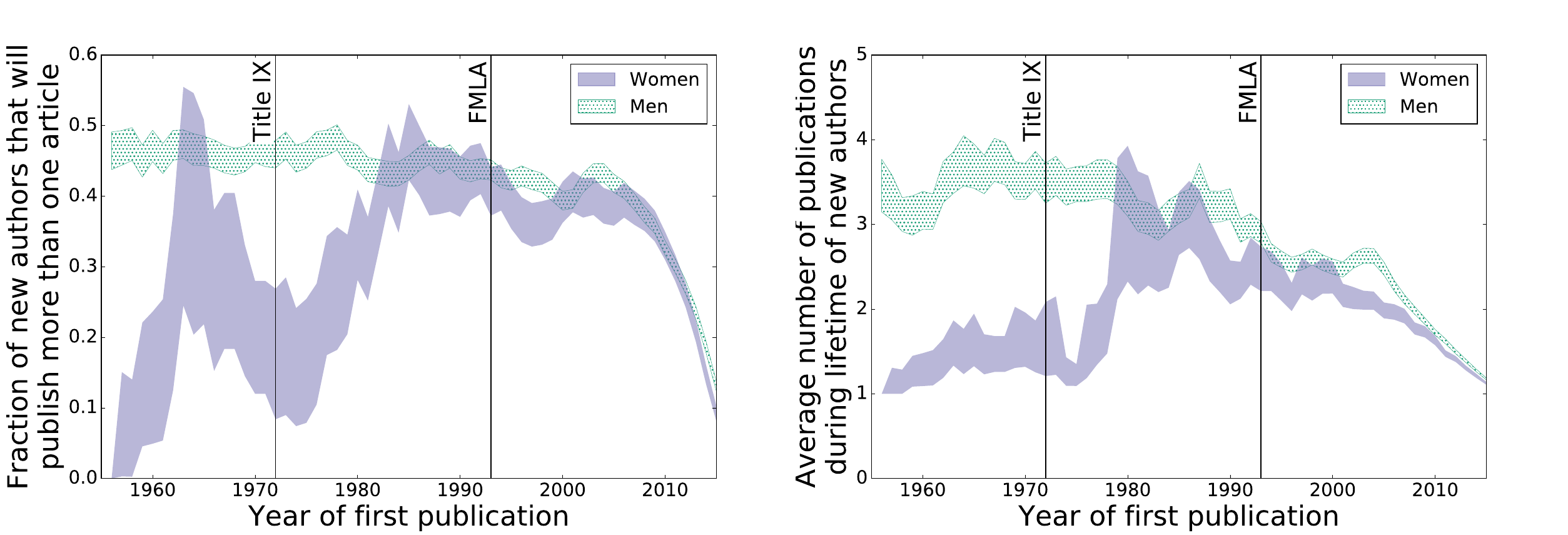}%
\caption{\csentence{In the later years, the fraction of new INFORMS women authors with more than one publication is similar to that of men, but \minorchange{their} average number of publications \minorchange{remains lower.}}
\textit{Left:}~Fraction of new authors that will publish again in INFORMS by 2016, \minorchange{as a function of their first year of publication}. 
\textit{Right:}~Average \minorchange{total} number of publications \minorchange{(as of 2016) by an author, also as a function of their first year of publication.  
A moving window was applied as in Figure~\ref{fig:authorship}. 
Shading} denotes $\pm1$ standard error of the mean (computed as in Figure~\ref{fig:authorship} for the left plot, and using $\mu \pm \sigma/n^{\sfrac{1\!}{2}}$, i.e., assuming a sum of independent Poisson variables, for the right \minorchange{plot). 
The} decrease in both quantities in later years is due to the fact that recent new authors have had less time to publish again. 
}
\label{fig:persistence}
\end{figure}

The fraction of authorships by women is consistently lower than the fraction of new women authors, reflecting the fact that women continue to have a lower average number of publications than men (Figure~\ref{fig:degreedistribution} and Movie~S1).
Moreover, the cumulative degree distributions further suggest that women who have many publications are disproportionately \newchange{rare.  
We quantified this effect by measuring the \mbox{power-law} exponent of the degree distributions, and found a significantly steeper slope for women (Figure~\ref{fig:powerlaw}).} 
\change{This claim is further supported by the results shown in Figure~\ref{fig:persistence}.}

\newpage
\section*{Two \minorchange{\mbox{diffusion-based}} measures for quantifying \minorchange{the importance\\ of a connection to the global structure of the network}}
\spaceaftersection

We \minorchange{now discuss the measures we used to quantify the importance of a connection between an author and publication to the overall connectivity of the network (as opposed to local characteristics such as degree, clustering, or other node/edge attributes). }  
\change{Our motivation for focusing on these particular measures is twofold: their use in graph algorithms relevant to the field of network science, and their direct relation with diffusion (arguably the simplest process that is sensitive to the global structure of a network). 
Additional global statistics of the INFORMS network are contained in Supplementary Information Table~\ref{table:summarystatistics}.}

 \change{The evolution of diffusion is governed by the graph Laplacian (defined as $L=D-A$, where $A$ is the adjacency matrix, and $D$ is the diagonal matrix of node degrees). }
 \minorchange{Indeed, many questions about the connectivity of a network (e.g., max-cut/min-flow problems~\cite{christiano2011electrical}, community detection~\cite{fiedler1973algebraic}) have efficient solutions that rely on the graph Laplacian~\cite{teng2010thelaplacian}. 
 In this section, we describe two measures of edge importance derived from the action of the Laplacian (more specifically, its pseudoinverse). }

\subsection*{\textbf{Effective resistance $\Omega$}}
\spaceaftersubsection

\minorchange{A} measure known \minorchange{as effective} resistance $\Omega$ naturally arises when quantifying the importance of an edge with respect to preserving the action of the graph \minorchange{Laplacian~\cite{spielman2011graph}. 
The} effective resistance $\Omega_e$ of an edge $e=(v_1,v_2)$ is defined as
\begin{equation}
    \Omega_e = b_e^\top L^\dagger b_e^{ },
\end{equation}
where $L^\dagger$ is the Laplacian pseudoinverse, and $b_e$ is the signed incidence (column) vector associated \minorchange{with} edge $e$, with nonzero entries for the two nodes adjacent to that edge:
\begin{align}
\left(b_e^{ }\right)_i^{ } &= \left\{
\!\!\!\begin{array}{rl}
+1&  i=v_1\\[-2pt]
- 1&  i=v_2\\[-2pt]
 0& \text{otherwise}
\end{array}
\right.
\end{align}

Aside from its algorithmic applications, $\Omega$ has a variety of intuitive interpretations.  
For example, it is\minorchange{:} the fraction of spanning trees that include this edge \cite{bollobas2001random}; the fraction of random walkers that use this edge during their stochastic transit between the two nodes \minorchange{joined by} this edge \cite{tetali1991random}; and, if one imagines \minorchange{the network as an electrical circuit where all the edges have unit resistance, it is the} voltage difference between the two nodes when passing a unit of current between them \cite{christiano2011electrical,chandra1989theelectrical}.

Hence, an edge with higher effective resistance is more important for diffusion between its adjacent nodes. 
In the context of authorship networks \minorchange{such as ours}, these connections often form ``bridges'' between \minorchange{communities}, whereas edges that are more redundant (i.e., have lower effective resistance) will tend to appear within \mbox{well-connected} \minorchange{groups.}

\subsection*{\textbf{Edge contraction importance $\Psi$}}
\spaceaftersubsection

\minorchange{However,} the  effective resistance \minorchange{measure} assigns \minorchange{its} maximal value (\mbox{$\Omega = 1$}) \minorchange{to} \textit{every} edge whose \minorchange{removal} would lead to a disconnection of the network, regardless of the sizes of the resulting components. 
This applies not only to edges that would disconnect large groups, but also to isolated edges and edges connecting a single node at the periphery of the network\minorchange{, which} are intuitively less important. \minorchange{ Hence, we also consider the ``contraction importance'' $\Psi$ \cite{bravohermsdorff2019aunifying} of edges, a recently proposed measure that is sensitive to these differences and also uses the graph Laplacian}. 

\begin{figure}[h!]
\includegraphics[width=0.9\linewidth]{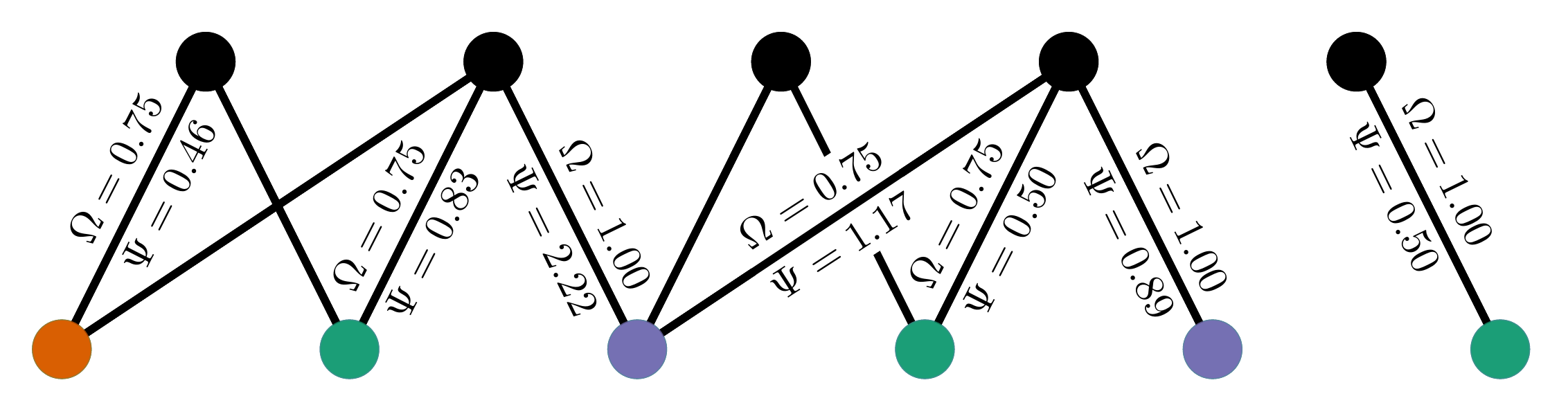}%
\caption{
\change{\csentence{Schematic illustrating the difference between effective resistance $\Omega$ and contraction importance $\Psi$ in a gendered bipartite authorship network.}}
This example network has the basic structure of the bipartite authorship network we studied: 
a collection of publications (represented by the \textbf{black} nodes) connected to their authors (represented by the nodes with colors corresponding to their assigned gender labels: ``\textcolor{womencolor}{\textbf{women}}'', ``\textcolor{mencolor}{\textbf{men}}'' and ``\textcolor{unknowncolor}{\textbf{unknown}}'').  
Effective resistance and contraction importance quantify the importance of an edge to diffusion in the network, albeit in different ways. 
The effective resistance $\Omega$ takes one of two values throughout the entire example network:~either $1.00$ (if its removal would disconnect nodes joined by that edge) or $0.75$ (if the edge participates in a $K_{2,2}$ substructure). 
In contrast, the contraction importance $\Psi$ is sensitive to the edge position relative to the rest of the network; an edge whose removal splits a network into two components with $n_A^{}$ and $n_B^{}$ nodes has importance \mbox{$\Psi = n_A^{} n_B^{}/n_{A\cup B}^{}$}.  
For example, an isolated edge has \mbox{$\Psi=0.50$} and an edge that connects to the giant component via one node has \mbox{$\Psi \rightarrow 1$}. 
In addition, edges that are more integral to the horizontal diffusion are given a higher $\Psi$. 
For instance, compare the edge with \mbox{$\Psi=1.17$} and the edge to its right with \mbox{$\Psi=0.50$}; the former is clearly more important for communicating between the left and right portions of the network, despite the fact that they both have \mbox{$\Omega=0.75$}.  
}
\label{fig:omegaandcontraction}
\end{figure}

The contraction importance reflects how much the dynamics of diffusion throughout the network would change if an author were \minorchange{\textit{merged}} with their publication. 
Specifically, it is the change in the Laplacian pseudoinverse incurred by this \minorchange{merge, as} \minorchange{quantified} by the Frobenius norm.
\minorchange{With} this measure, an author who is a bridge between two communities \minorchange{is considered more important than an author who is only connected to the periphery of the network (see Figure~\ref{fig:omegaandcontraction}). 
The} contraction importance $\Psi$ of an edge $e$ is defined as:
\begin{equation}
    \Psi_e^{ } = \frac{b_e^\top L^\dagger L^\dagger b_e^{ }}{b_e^\top L^\dagger b_e^{ }}
\end{equation}

\minorchange{To summarize, while} both $\Omega$ and $\Psi$ give a measure of the importance of an edge to the overall connectivity of the network, there are some notable differences, particularly in the way they treat edges whose removal would disconnect a component of the \minorchange{network.} 
\minorchange{$\Omega$} considers how \minorchange{much the} diffusion between \minorchange{a given} author and \minorchange{one of their} publications would be \minorchange{reduced} if \minorchange{this connection} were \minorchange{to be} deleted (i.e., removing the author from \minorchange{this} publication). 
\minorchange{Thus, if this connection} is the only path from this author to this publication, \minorchange{it would be considered maximally important (\mbox{$\Omega = 1$}).}
\minorchange{In contrast, $\Psi$ measures how much the contraction of an author--publication connection (i.e., considering this author and publication as the same entity) would \minorchange{alter} diffusion throughout the \minorchange{\textit{entire}} network. 
\minorchange{Therefore, contraction importance does not treat all edges with \mbox{$\Omega = 1$} equally, instead assigning smaller values to those that are less important for \textit{large-scale} diffusion throughout the network (indeed the case for most edges with $\Omega = 1$).}
}  

\newpage
\subsection*{\textbf{\minorchange{Effective resistance $\Omega$ and contraction importance $\Psi$ distributions\\ in the INFORMS network}}}
\begin{figure}[h!]
\includegraphics[width=0.98\linewidth]{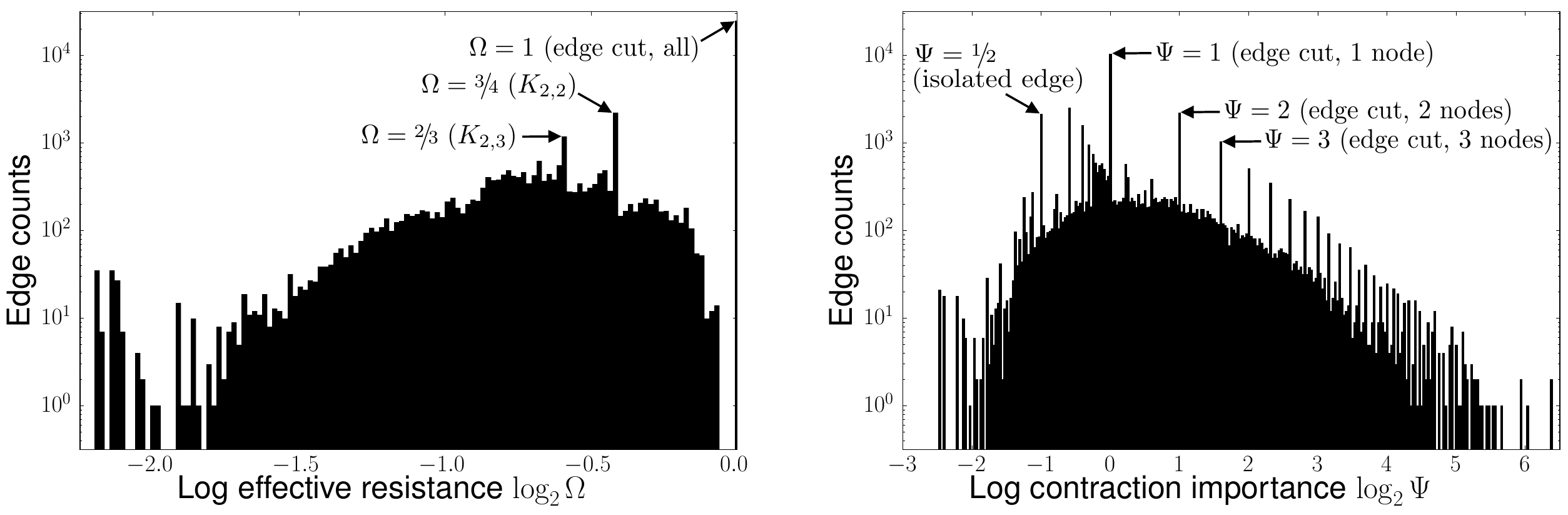}
\caption{\csentence{Distributions of log effective resistance (\mbox{$\log_2 \Omega$}) and log contraction importance (\mbox{$\log_2 \!\Psi$}) provide information about the connectivity properties of the network.}
Peaks in the effective resistance at $\sfrac{3\!}{4}$ and $\sfrac{2\!}{3}$, correspond to the complete bipartite subgraphs $K_{2,2}$ and $K_{2,3}$ (i.e., two authors/publications connected to two or three publications/authors), respectively. 
Peaks in the contraction importance at \mbox{$n \in \mathbb{N}$}, correspond to edges whose removal would separate $n$ nodes from the bulk of the network.  
\minorchange{The counts of such edges in these integer peaks appears to approximately follow a power law (with exponent \mbox{$\gamma\sim2$}).}  
}
\label{fig:edgeimportancemeasures}
\end{figure}

\spaceaftersubsection
\minorchange{When} comparing the relative importance of \minorchange{different} edges, one should consider their ratio\minorchange{. 
Hence, we take the log of these measures} before computing means and other statistics.  
Figure~\ref{fig:edgeimportancemeasures} displays the histogram of both measures in the entire cumulative INFORMS network, \minorchange{illustrating} how they provide relevant information about the connectivity of the network. For example, the contraction importance exhibits peaks at \mbox{$n \in \mathbb{N}$}, corresponding to edges whose removal would separate $n$ nodes from the bulk of the network, and the effective resistance exhibits peaks at $\sfrac{2\!}{3}$ and $\sfrac{3\!}{4}$, corresponding to the complete bipartite subgraphs $K_{2,3}$ and $K_{2,2}$ (i.e., two authors/publications connected to three or two publications/authors). 

Both measures have similarly shaped distributions when conditioned on the genders.  
However, the average log contraction importance of authorships by women is lower than for authorships by men, while the average log effective resistance of authorships by women is higher (see Figure~\ref{fig:edgeimportancesinthemodel}). 
In order to provide a meaningful comparison of these measures between the genders, we \minorchange{constructed} a null model that explicitly replicates the yearly degree distributions for publications, women authors, and men authors, but assigns author--publication connections irrespective of gender.

\subsection*{\textbf{A degree-preserving temporal and geometric null model\\with emergent communities}}
\spaceaftersection

Gender differences in global importance measures could be attributable to gender differences in local statistics.  
For example, edges connecting nodes with higher degrees tend to have lower effective resistance, thus the sole fact that men have a higher number of publications could directly lead to \change{the observed} difference between \change{the} effective resistance of authorships by women and men. 
Therefore, in order to determine if differences in collaboration patterns require an explanation beyond \change{basic} local statistics, it is crucial to have a null model that replicates the relevant local properties of the original network. 
Such a null model can then be used to determine \change{whether} other \change{properties} of interest are simply consequences of \change{these local statistics}.

In this section, we describe a novel degree-preserving temporal and geometric null model with emergent communities, which we believe could be of independent interest. 
\change{The model employs} a combination of geometric embedding to mimic \change{clustering due to node} similarity \cite{krioukov2016clustering}, and a self-reinforcing \mbox{node-placement} \change{mechanism} to \change{encourage community formation} \cite{zuev2015emergence}.

The model explicitly replicates the observed yearly degree distributions for publications, as well as the \change{gender and} publication histories of each individual author \change{(i.e., their number of publications each year)}. These are taken from the data as fixed inputs to the model. As we are interested in the effect of gender on collaboration patterns, our null model is blind to gender when assigning author--publication connections.

Aside from these inputs, the model has two free parameters: $D$, the dimension of the embedding space, and \change{$n_{\textrm{nei}}$, a parameter controlling the propensity for clustering. } 
We set these parameters so as to best match the data in other relevant aspects, such as the component size distribution over time.

Specifically, for each year, the model performs the following steps: 

\begin{enumerate}
\item Choose a length scale $\lambda$ for this year, such that there will be $n_{\textrm{nei}}$ expected number of authors within a ball of radius $\lambda$, i.e., $\lambda = \left( \frac{n_{\textrm{nei}}}{n_{\textrm{tot}}} \frac{\Gamma(\sfrac{D\!}{\!\!\;2}+1)}{\pi^{\sfrac{D\!}{\!\!\;2}}} \right)^{\!\!\sfrac{1\!}{\!\!\;D}}$, where $n_{\textrm{tot}}$ is the number of author nodes already placed (from previous years). 
    
\item Add the new author nodes from this year (simultaneously) at locations \change{$x_a$} in \change{a} $D$-dimensional \change{unit} torus, with probability proportional to the ``attractiveness'' of that location:
\begin{equation}
    p(x_a) \propto \sum_{i=1}^{n_{\textrm{tot}}} \exp{\left(-\frac{|x_a-x_i|^2}{2\lambda^2}\right)},
\end{equation}
where $x_i$ are the locations of the existing $n_{\textrm{tot}}$ author nodes from previous years.
\item Add a ``\mbox{half-edge}'' to each of the author nodes, for each publication they authored this year.
\item Sequentially add the publication nodes from this year, in order of \change{decreasing number of authors}.  
The probability of placing a publication at a location $x_p$ is 
\begin{equation}
    p(x_p) \propto \prod_{i=1}^{d_p} \exp{\left(-\frac{|x_p-x_i|^2}{2\lambda^2}\right)},
\end{equation}
where $d_p$ is the number of authors on the publication, and $x_i$ are the locations of the nearest $d_p$ authors with available \mbox{half-edges}.  
\item The publication then connects to each of its requisite number of authors with probability
\begin{equation}
    p(x_p \leftrightarrow x_a) \propto \exp{\left(-\frac{|x_p-x_a|^2}{2\lambda^2}\right)},
\end{equation}
where $x_p$ and $x_a$ are the positions of the publication and an author with an unused \mbox{half-edge}, respectively. 
\item Repeat steps \mbox{4--5} until all publications haven placed (and author \mbox{half-edges} used). 
\end{enumerate}

In the network created by this algorithm, every woman and man author has exactly the same number of publications \change{for each year} as they do in the INFORMS network. 
However, importantly, the mechanism by which the null model decides which publication to give to an author does not consider gender\change{, and thus the resulting networks can be used as a null model to study gender differences in global statistics}. 

\subsection*{\textbf{The null model mimics emergent network properties}}
\spaceaftersubsection

\begin{figure}[h!]
\includegraphics[width=0.97\linewidth]{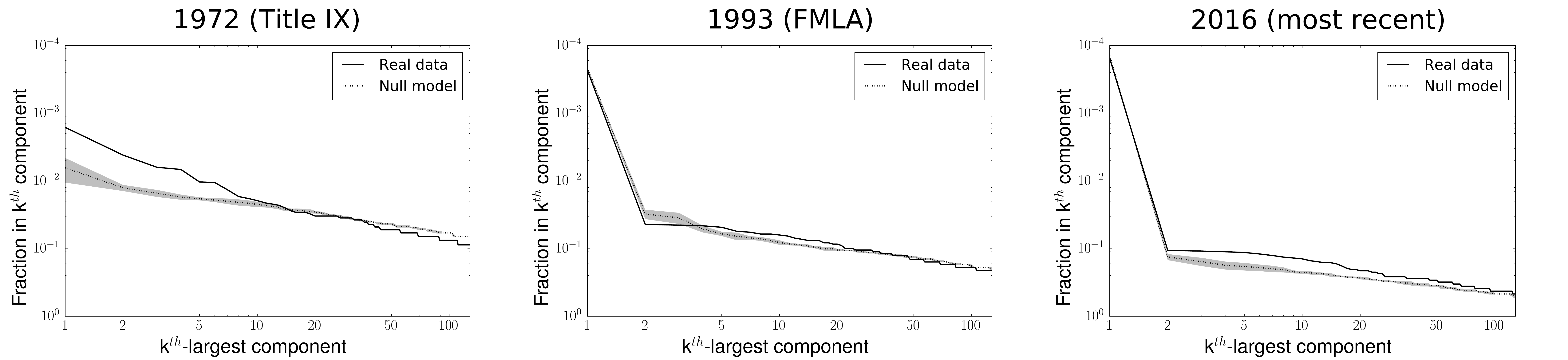}%
\caption{
\minorchange{\csentence{The null model qualitatively mimics the emergence and size of the giant component.} } 
Plots show the fraction of authors in each of the 128 largest components in the cumulative network up to \minorchange{the indicated year.} 
Solid black curve denotes the \mbox{INFORMS} network, and the shading denotes the range ($\pm1$ standard deviation) of a typical run of the null model (with parameters \mbox{$D=4$} and \mbox{$n_{\textrm{nei}}=1$}) about the dotted null model mean. Statistics were computed by running 8 instances of the null model. 
}
\label{fig:components}  
\end{figure}

Aside from the explicitly matched temporal degree distributions, our null model is able to capture other relevant network features (see Figure~\ref{fig:components}). Namely, certain settings of the free parameters (e.g., \mbox{$D\gtrsim4$}) allow the null model to replicate the initial appearance of the giant component, while others (e.g.,  \mbox{$D\lesssim4$}) match better the behavior in later years. 

\subsection*{\textbf{Correlations between author gender and the role of their authorships\\in the global network structure}}
\spaceaftersubsection

\begin{figure}[h!]
\includegraphics[width=0.97\linewidth]{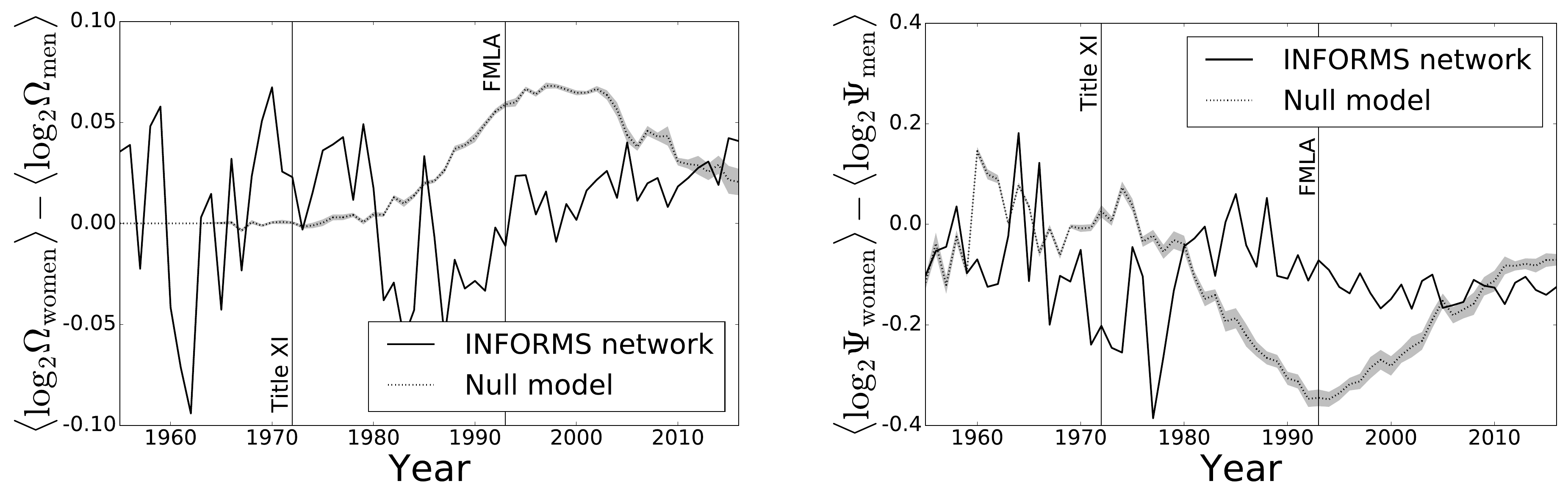}
\caption{\minorchange{\csentence{Evolution of measures of authorship importance indicates structural differences between the genders that have recently tapered off.} }
Gender differences (women minus men) over time \minorchange{in average edge importance measures (effective resistance, $\log_2 \Omega$, left; and contraction importance, $\log_2 \Psi$, right)}. 
The solid curve is the difference in the INFORMS dataset, and the dotted curve in the null model, which replicates the yearly degree distributions for publications, women authors, and men authors, but assigns author-publication connections irrespective of gender.
\minorchange{The shading denotes the range ($\pm1$ standard deviation) of a typical run of the null model (with parameters \mbox{$D=4$} and \mbox{$n_{\textrm{nei}}=1$}) about the dotted null model mean (statistics were computed by running 8 instances of the null model).}
}
\label{fig:edgeimportancesinthemodel}
\end{figure}

We used our null model to evaluate the effect of gender in determining the importance of an authorship to the global network structure. 
For every year, we calculated the difference in \minorchange{average edge importance measures ($\log_2 \Omega$ and $\log_2 \Psi$)} between the genders for the INFORMS \minorchange{network} and for multiple simulations of the null model (all with \mbox{$D=4$} and \mbox{$n_{\textrm{nei}}=1$}). 
As \minorchange{shown} in Figure~\ref{fig:edgeimportancesinthemodel}, the data \change{and null model deviate considerably.}  

\change{Until around \mbox{1980}, the null model predicted no difference between the effective resistance $\Omega$ of authorships by women and men. 
This is likely due to the prevalence of \mbox{author--publication} connections with \mbox{$\Omega=1$}. 
In fact, until 1966, there were only 21 women in the dataset, and all had only one publication; this means that all of their effective resistances are precisely $1$ in both the data and the null model. 
Moreover, as the null model predicted no difference during this period between the effective resistance of authorships by women and men, authorships by men similarly have effective resistance close to $1$. 
This is in contrast to the easily-discernible clustering present in the earlier years of the actual INFORMS network -- our choice of parameters for the null model resulted in networks containing more tree-like components (for which all edges have \mbox{$\Omega=1$}) during this period.  
We chose these parameters in order to qualitatively match the bulk of the available data, namely the emergence (\mbox{$\sim$1980}) and size of the giant component. 
While our model captured this feature of the data, none of the parameter settings we tried allowed it to capture the clustering present before $\sim$1980. 
Therefore, it is difficult to interpret differences between the null-model and the actual data during this early period. 
The average contraction importance does, however, show gender differences in the null model during this period (Figure~\ref{fig:edgeimportancesinthemodel}, right), highlighting the higher sensitivity of this measure of edge importance. 
}

\change{From around 1980 until around 2005, the null model begins to predict gender differences in both $\Omega$ and $\Psi$.  
This is a particular relevant period as $\sim$1980 coincides with the onset the of a consistent increase in women participating in INFORMS (see Figure~\ref{fig:authorship}) and the emergence of the giant component (see Supplementary Video 1), and we choose the parameters of the null model to best replicate the component size distribution during this period (see Figure~\ref{fig:components}). 
Interestingly, the two measures deviate in opposite directions: the null model predicts that women should have a higher effective resistance and lower contraction importance than men. 
Neither of these changes were observed in the actual data, which instead display gender differences closer to zero for both $\Omega$ and $\Psi$. 
This suggests that the gender differences in global statistics found in the INFORMS network during this period cannot be explained using only local statistics. 
Finally, from $\sim 2005$, the predicted gender differences are closer to those observed in the actual data, where authorships by women have effective resistance slightly higher and contraction importance slightly lower than authorships by men. 
This suggests that the more recent gender differences in global statistics can be explained mostly by differences in local statistics, such as the underrepresentation of women in the network. 
}

\section*{Discussion}
\spaceaftersection

\minorchange{In this work, we} investigated the relevance of gender in scientific \minorchange{publication patterns} by analyzing a \minorchange{temporal bipartite network between authors and publications in the INFORMS journals. }
Our study provides two methodological contributions:
\minorchange{\text{1)} We develop a simple} temporal geometric null model \minorchange{(with two free parameters)} that encourages emergent communities (a hallmark feature of real-world networks)\minorchange{;  and} \text{2)} we present a case study for applying two measures of edge importance related to diffusion throughout the network\minorchange{, namely effective resistance $\Omega$ and contraction importance $\Psi$. }

\change{
While conventional measures of importance/centrality (e.g., edge betweenness \cite{girvan2002community}) may give qualitatively similar results, the measures we chose are particularly relevant due to their use in the analysis and implementation of a variety of graph algorithms. 
For example, they serve as a measure of edge importance in several graph reduction algorithms \cite{spielman2011graph, fung2019general}, which are often used as primitives in other efficient algorithms for massive networks.  
As the field of applied network science often deals with such structures, it is relevant to investigate how such algorithms might interact with datasets containing metadata.
}

\minorchange{At the level of local statistics, we} found that both the fractions of new women authors and of authorships by women in the INFORMS network have been increasing steadily since $\sim$1980 (see Figure~\ref{fig:authorship}). Before then, the fraction of new women authors hovered around less than 3\%. 
While multiple factors may have contributed to the sudden change in derivative around 1980, it is not unreasonable to hypothesize that the introduction of Title IX legislation in the United States in 1972 played a role that is still unfolding, and such quantification of long-term effects could serve as a good argument for similar policies.

Despite the continued increase in the fraction of new women authors in INFORMS, more than four decades after Title IX, women \minorchange{still comprise less} than $25\%$ of the network (Figure~\ref{fig:authorship}).
\change{Moreover, women are disproportionately underrepresented among authors with many  publications (Figures~\ref{fig:degreedistribution} and~\ref{fig:powerlaw}).} 
While this could be due to a variety of factors, we draw attention to gender biases in the peer-review process and visibility of researchers in the community (e.g., invitations to present work at conferences and colloquia \cite{nature2016women}). 
These issues have been the focus of several recent studies. 
For example, Murray et al.~\cite{murray2018gender} studied the review process in the journal eLife, and found that reviewers appear to favor \minorchange{authors with demographic characteristics (gender and nationality) similar to their own.  }
In analyzing data from a longitudinal experiment by the Canadian Institutes of Health Research, Witteman et al.~\cite{witteman2019gender} found that women were less likely to be awarded a grant when the review focus was on ``the principal investigator'' as compared to \minorchange{``the proposed science''}.  
Nittrouer et al.~\cite{nittrouer2017gender} found that men were more likely than women to be colloquium speakers at top US universities (even when controlling for speaker rank and the gender ratio in the field, and despite \minorchange{men and women declining invitations at similar rates}. 
Due to enhanced scrutiny and attention to these issues, these trends have been changing more recently. 
For example, BiasWatchNeuro.com~\cite{bwn}, a website dedicated to monitoring gender representation of invited speakers at conferences in the field of neuroscience, has documented a steady increase in the rate of women invited to present at conferences, with this rate now approaching a conservative estimate of the base rate of women in the field ($\sim30\%$, \cite{biaswatchneuro_2018}). 

\minorchange{Indeed, much} effort is being invested in methods for mitigating these biases, such as compiling online lists of women researchers to facilitate \minorchange{their} invitation \minorchange{as} conference speakers and nomination for prizes \cite{request, academianet, winrepo}; making data on the gender balance in conferences and panels more visible online \cite{bwn}; and encouraging journals to adopt a policy of double-blind reviews, which has been showed to reduce biases (such as increasing representation of women authors \cite{budden2008double}). 
Such efforts might help improve scientific productivity by increasing gender heterogeneity in the scientific workforce~\cite{campbell2013gender}. 
For a thorough review on the issue of gender bias in science (with a focus on neuroscience) and recommended efforts to mitigate it, see Schrouff et al.~\cite{schrouff2019gender}.

\change{By accurately quantifying the nature of systematic asymmetries, one can more precisely inform policies intended to remedy them~\cite{fagan2018assessing, luke2016forging, okamoto2014scientific}. }
\minorchange{For example, simulations of STEM faculty retention in a US university \cite{thomas2015gender} suggest that in order for gender parity to be reached in those data, the higher rate of women resigning must be addressed.
Follow-up studies investigating the reasons behind the higher resignation rate of women in STEM fields could be particularly helpful in informing policies targeting this attrition.  
Indeed, a}  recent study~\cite{cech2019changing} analyzing a longitudinal national survey of US STEM professionals, suggests that \minorchange{this} higher rate \minorchange{might} be due to \minorchange{women leaving their} full-time STEM jobs upon becoming parents for the first time ($\sim43\%$ resignation of women vs.~$\sim23\%$ of men). 
Additionally, it was found that, among those that continued to hold full-time jobs, parents were less likely to remain in their STEM jobs than their child-less peers. 
This suggests that policies to render work in STEM fields more compatible with caregiving may be critical to increase gender \minorchange{diversity. } 
\change{Likewise, network science analysis such as ours could help inform diversity efforts by indicating potential connections that are more likely to decrease gender disparities, for example, by funding research with \mbox{author--publication} connections that are diverse and important to the connectivity of the network. 
This type of analysis could also help evaluate the global impact of such efforts. 
For instance, future work could focus on trends that might correlate with more recent diversity initiatives in the INFORMS society (e.g., in 2006 and 2017).  
However, as the time-scale of our dataset is on the order of a typical career length, results for the last decade or so of our data are still unfolding; thus, it is presently difficult to measure the effect of policy changes during this time. }

To shed light on where to focus such efforts, future \minorchange{work could} focus on correlations between gender differences and various subcommunities. 
For example, \minorchange{in the current INFORMS dataset}, some journals have a \minorchange{relatively} higher rate of women authors (e.g., $\sim28\%$ in \textit{Organization Science}), while others have a \minorchange{much lower rate} (e.g., $\sim8\%$ in \textit{Mathematics of Operations Research}; see supplementary Figure~\ref{fig:genderperjounal}).
Another interesting research avenue is to understand how particular collaboration patterns and homophily correlate with academic success (as measured, for instance, by number of future publications). 
The period of 1980--2005 in the INFORMS network is especially interesting in this regard, given indications that collaborative and publication patterns were different between genders in those years in ways that are not explained by local statistics. Our results suggest that, on average, women and men had more similar effective resistance and contraction importance in those years than predicted by the local statistics \change{incorporated in our} null model. 
This \change{discrepancy between the data and the null model could} indicate that women entering the INFORMS network had similar network roles as men. 
Another possibility is that women and men formed somewhat separate collaborative networks with similar statistics.
Future work could investigate whether these, or other patterns, are dominant in the network.

\change{It is important to recognize the limitations of our methodology.  
For instance, our data cleaning methods are likely biased: Genderize.io (the software we used to classify the genders of the authors) has a higher misclassification rate for \mbox{non-Western} names~\cite{santamaria2018comparison}. 
Moreover, as first names in many contemporary Western cultures are frequently indicative of only two genders, the API reduces the multidimensional continuum of genders to a single binary variable. }
\change{In addition, the data selected for manual gender classification (i.e., all authors with more than six publications) are also likely correlated with gender,}\footnote{\footnotesize{\change{Of authors with 6 or fewer publications, $\sim 13\%$ were classified as women, $\sim 69\%$ as men and $\sim 17\%$ as unknown, whereas the composition of those with more than 6 publications was $\sim 7\%$ women and $\sim 93\%$ men.}}} \change{biasing the ratio of known author genders (as women have disproportionately fewer publications than men).} 
\minorchange{It is also} important to acknowledge that our simple and physically-motivated null model does not replicate all of the relevant gender-ambivalent network properties. 
For example, we did not precisely match the component size distributions throughout the development of the network; lower dimensional embeddings better matched the earlier distributions, and higher dimensional embeddings better matched the later distributions.  
A time-varying $n_{\textrm{nei}}$ will likely describe the data better, and thus provide a better comparison for evaluating the roles of women and men in the network over time, in particular, in early years of the network. 
However, from a practical point of view, those early years may be less informative regarding the effectiveness of current policies, and suggestions for further interventions to mitigate gender biases.

Overall, our results indicate that the INFORMS \change{society remains far from gender parity in many important local statistics. 
However, since $\sim$1980 these differences have been steadily decreasing, and recent ($\gtrsim$2005) global statistics are more in agreement with our null model, indicating some progress. 
More generally, we hope that building a quantitative understanding of gender publication and collaboration patterns in academia will ultimately help accelerate the path towards gender equality by bringing awareness to the issue and informing future studies and policies. 
}

\newpage
\section*{Supplementary information}
\spaceaftersection

\subsection*{\textbf{Video of evolution of cumulative degree distributions by gender.}}
\spaceaftersubsection
\change{The file with this video is provided in the Supplementary Material. }

\subsection*{\textbf{\change{Additional descriptive statistics.}}}
\spaceaftersubsection

\change{Table~\ref{table:summarystatistics}  displays additional descriptive statistics of the INFORMS dataset. 
We defined the time between publications for each author as $(\text{final year}-\text{initial year})/(\text{\# publications}-1)$, and computed the averages including only authors with more than one publication ($\sim$35\% of authors; $\sim$31\% of women, $\sim$39\% of men and $\sim$23\% of unknown gender). }
\begin{table}[h!]
\begin{tabular}{l!{\vrule width1pt}c!{\color{womencolor}{\vrule width1pt}}c!{\color{mencolor}{\vrule width1pt}}c!{\color{unknowncolor}{\vrule width1pt}}}
& \vphantom{\Big)}\!\textcolor{womencolor}{\textbf{women authors}}\! & \vphantom{\Big)}\!\textcolor{mencolor}{\textbf{men authors}}\! & \vphantom{\Big)}\!\textcolor{unknowncolor}{\textbf{unknown authors}}\!\\
\Xhline{2\arrayrulewidth}
\vphantom{\large)}num.~of authors & $2997$ & $16179$ & $3735$ \\
\hline
\vphantom{\large)}avg.~num.~of~publications per author & $1.82$ & $2.47$ & $1.36$ \\
\hline
\vphantom{\large)}avg.~num.~of co-authors & $3.29$ & $3.81$ & $2.81$ \\
\hline
\vphantom{\large)}avg.~num.~of journals published in & $1.27$ & $1.42$ & $1.16$ \\
\hline
\vphantom{\large)}percentage~of authorships & $0.11$ & $0.79$ & $0.10$ \\
\hline
\vphantom{\large)}percentage~of $1^{\text{st}}$ authorships  & $0.10$ & $0.80$ & $0.10$ \\
\hline
\vphantom{\large)}percentage~of solo authorships & $0.06$ & $0.86$ & $0.08$ \\
\hline
\vphantom{\large)}avg.~time between publications & $4.14$ & $4.23$ & $4.11$ \\
\hline
\vphantom{\large)}median time between publications & $3.00$ & $3.00$ & $3.00$ \\
\hline
\vphantom{\large)}median time of first publication & $2007$ & $1999$ & $1994$ \\
\hline
\vphantom{\large)}num.~publications of top 5$\%$ (range) & $5$ up to $32$ & $8$ up to $90$ & $3$ up to $6$ \\
\Xhline{2\arrayrulewidth}
\vphantom{\large)}avg.~log effective resistance $\log_2 \Omega$ & $-0.27$ & $-0.31$ & $-0.31$ \\ 
\hline
\vphantom{\large)}avg.~log contraction importance $\log_2 \Psi$ & $0.21$ & $0.33$ & $0.31$ \\ 
\hline
\vphantom{\large)}avg.~normalized edge betweenness & $9.96\cdot10^{-5}$ & $12.7\cdot10^{-5}$ & $3.97\cdot10^{-5}$ \\ 
\hline
\end{tabular}
\vspace{10pt}
\caption{\change{ \textbf{Descriptive statistics of the entire cumulative INFORMS network through 2016.}} }
\label{table:summarystatistics}
\end{table} 

\newpage
\subsection*{\textbf{Authors gender in individual INFORMS journals.}}
\spaceaftersubsection

\begin{figure}[ht]
\includegraphics[width=0.95\linewidth]{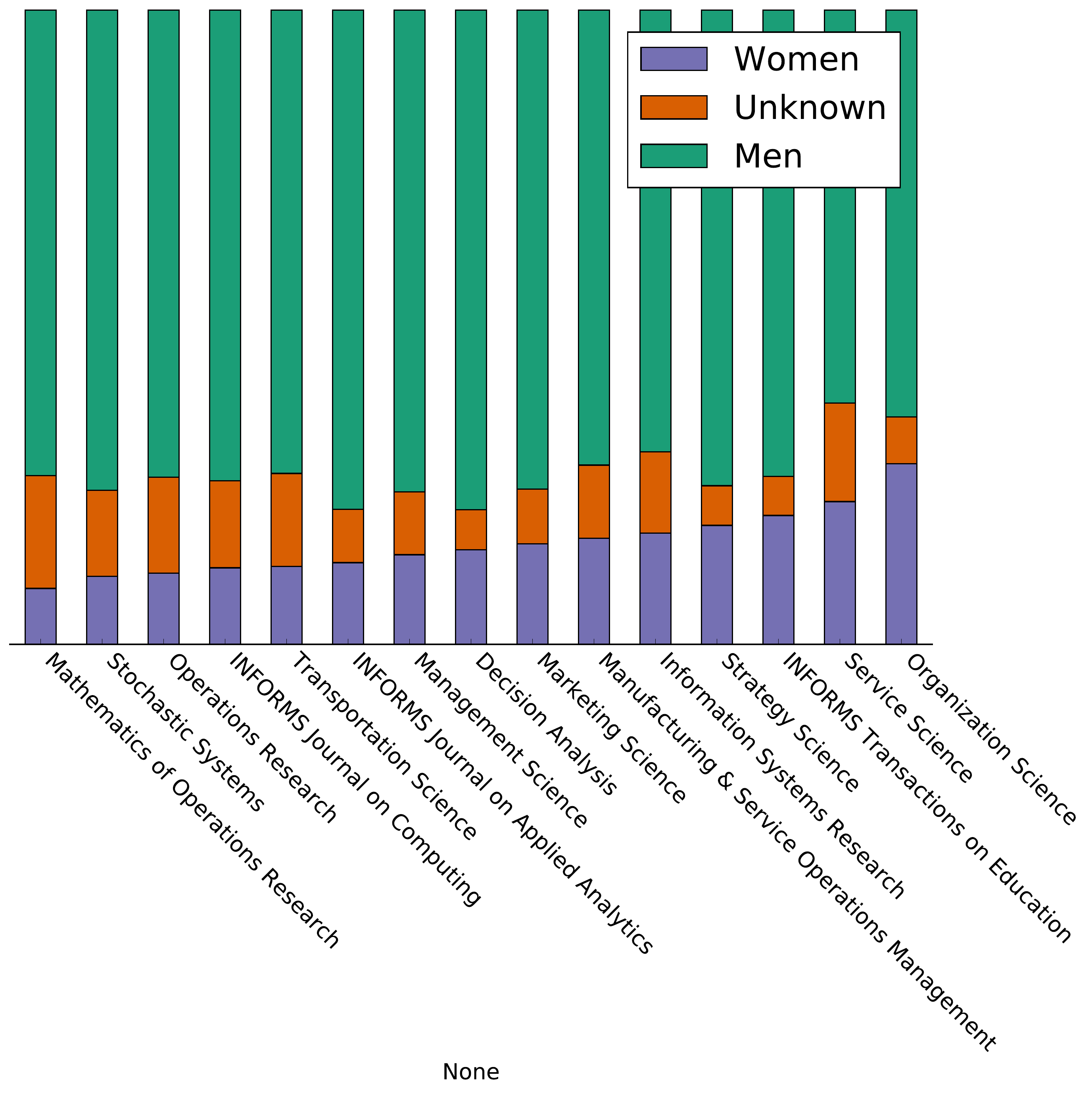}
\caption{\minorchange{\csentence{Fraction of genders of unique authors within each INFORMS journal.} }
Note that the journal \textit{Management Technology} merged with the journal \textit{Management Science} in 1965. 
Here, we included authorships in \textit{Management Technology} as authorships in \textit{Management Science}.
}
\label{fig:genderperjounal}
\end{figure}



\section*{List of abbreviations}
\spaceaftersection

INFORMS: Institute for Operations Research and the Management Sciences; 
STEM: science, technology, engineering, and mathematics; 
ORSA: the Operations Research Society of America;
TIMS: the Institute of Management Sciences;
Title IX: Title IX of the Education Amendments Act of 1972;
FMLA: The Family and Medical Leave Act of 1993

\section*{Declarations}
\spaceaftersection

\subsection*{\textbf{Competing interests}}
\spaceaftersubsection
The authors declare that they have no competing interests.

\subsection*{\textbf{Funding}}
\spaceaftersubsection
This work was funded in part by a diversity grant from the Simons Foundation (VF,YN) and internal funds from Princeton University (GBH,YN).

\subsection*{\textbf{Author's contributions}}
\spaceaftersubsection
ER, JM, and MEH conceived the overarching question and constructed the dataset. 
ER, GBH, MEH and VF performed data cleaning and validation. GBH and LMG designed and performed the analysis and discussed the results with ER, MEH, VF and YN. GBH, LMG, VF, and YN wrote the manuscript with input from ER and MEH. 
All authors read and approved the final manuscript.

\subsection*{\textbf{Availability of data and material}}
\spaceaftersubsection
Data are available from the authors upon reasonable request and with permission of INFORMS.

  
  

\bibliographystyle{bmc-mathphys} 
\bibliography{bmc_article}      

\end{document}